# 64-APSK Constellation and Mapping Optimization for Satellite Broadcasting using Genetic Algorithms

Matteo Anedda, Alessio Meloni and Maurizio Murroni
{matteo.anedda},{alessio.meloni},{murroni}@diee.unica.it
DIEE, Dept. Electrical and Electronic Engineering
University of Cagliari, Piazza d'Armi, 09123, Cagliari, Italy

*Abstract* — **DVB-S2 and DVB-SH satellite broadcasting standards currently deploy 16- and 32- APSK modulation using the consultative committee for space data systems (CCSDS) mapping. Such standards also include hierarchical modulation as a mean to provide unequal error protection in highly variable channels over satellite. Foreseeing the increasing need for higher data rates, this paper tackles the optimization of 64-APSK constellations to minimize the mean square error between the original and received symbol. Optimization is performed according to the sensitivity of the data to the channel errors, by means of genetic algorithms, a well-known technique currently used in a variety of application domains, when close form solutions are impractical. Test results show that through non-uniform constellation and asymmetric symbol mapping, it is possible to significantly reduce the distortion while preserving bandwidth efficiency. Tests performed on real signals based on perceptual quality measurements allow validating the proposed scheme against conventional 64-APSK constellations and CCSDS mapping.**

## I. INTRODUCTION

AMPLITUDE Phase Shift Keying (APSK) modulation with pre- and post- compensation schemes is deployed in satellite broadcasting [1]-[3] for its power and spectral efficiency over nonlinear satellite channels. Nevertheless, for multimedia broadcasting applications, further improvements by means of non-uniform constellations could be obtained [4]. As a matter of fact, multimedia streams employed in digital broadcasting are hierarchical by nature, so that bits associated with transmitted symbols present different error sensitivities. In particular, faults on most significant bits (MSBs) affect the transmission more than errors on the least significant bits (LSBs). For this reason, in the last years several works tackling the problem of guaranteeing scalable quality in satellite communications through hierarchical modulation have been presented [5]-[9] and integrated in standards such as DVB-S2 and DVB-SH. A hierarchical modulation carries two separate and independent bit streams. The primary (i.e., coarse) bit stream and the secondary (i.e., fine) stream. The former is intended for users with poor channel quality, whereas the latter stream refines the first one, but requires a larger signal-to-noise ratio (SNR) to be decoded error-free. Hierarchical modulation can be used effectively to upgrade a digital broadcast system in response to both the demand for higher bit rate that is made possible due to advances in technology and coding algorithm development, and the need to be backward compatible to the already deployed old receivers.

In hierarchical modulation fine and coarse streams present different sensitivity to channel error and different techniques can be implemented to assure robustness. Channel coding techniques (i.e., FEC coding) to implement unequal error protection (UEP) have been studied in [10] for quadrature amplitude modulation (QAM) and in [11] for APSK, even though this introduces overhead and reduces bandwidth efficiency, which is a critical issue for satellite applications [12], [13].

In [14] modulation with unequal power allocation (MUPA) was proposed as a mean to improve the performance of conventional modulation schemes. In fact, saving bandwidth is verified in case of digital wireless communication systems which do not include channel coding for some reason. MUPA achieves UEP hierarchically distributing the available budget power over the symbols according to their sensitivity to channel errors, whereas the average transmission power per symbol remains unchanged. The resulting quality on received data was measured by means of the mean square error (MSE) between the original and decoded symbol. By increasing the robustness of MSBs MUPA reduces the average distortion (i.e., MSE) between transmitted and decoded data and the quality achieved at receiver side is improved without any increase of transmission bandwidth. MUPA performance evaluation does not consider conventional bit error rate (BER) measurements, because the final goal is to achieve minimal distortion among symbols which are considered as composed by bits with different importance and which are therefore unequally protected by an opportune optimized power distribution process [14].

The MUPA concept was applied to APSK constellations for satellite applications in [15] and [16] where 16 and 32-APSK modulations have been treated in order to achieve UEP through asymmetric layout of the constellation symbols. The optimization problem (OP) aimed at selecting the opportune radius of the constellation circles and the phase of each symbol which minimize MSE (i.e., the average distortion) between transmitted and decoded data. Due to the complex nature of the OP which does not allow close form solution, genetic algorithms (GA) [17] have been deployed. GA is a search technique used in many fields to solve complex problems which do not allow analytical derivation

[18], [19]. GA fall within the class of gradient search techniques which need opportune setup to avoid sub optimal solutions due to the presence of local minima. In this work to set up the GA evolution the result of a previous study performed in [16] which analyzed the influence of the GA parameters on the accuracy and convergence performance of the algorithm has been considered.

The novelty of this work consists in studying the case of 64-APSK based on the results obtained in [15] and [16] foreseeing near future deployment of high order modulations in satellite broadcasting: new mappings derived starting from the one released by the consultative committee for space data systems (CCSDS) [20] and novel non-uniform constellations are proposed and compared with standard 64-APSK schemes. The outcome of this paper demonstrates that the use of such modified mapping and non-uniform 64-APSK constellations through GA optimization yields to a reduction of the perceived distortion at the receiver. In fact, GA optimization discriminates the most significant piece of information in the modulated symbol obtaining reduced MSE for the proposed system with respect to the state-of-the-art conventional 64-APSK with CCSDS mapping. Correspondence between MSE and subjective perception has been validated by test performed on real signals. Digital audio and image signals and have been generated and transmitted both with the proposed optimized constellation and with conventional 64-APSK. In case of audio, performance evaluation has been measured by means of the perceptual evaluation of audio quality (PEAQ) algorithm [21]. PEAQ is an objective measurement technique recommended by the International Telecommunication Union – Telecommunication Standardization Bureau (ITU-T), which evaluates the quality of an audio signal by a single number, called objective difference grade (ODG), which varies within a range [-4÷0], with 0 the highest quality score. PEAQ has proven to achieve higher performance than conventional metrics based on MSE on the evaluation of the performance of the conventional audio codecs [22].

In case of image transmission, direct subjective evaluation is given to the reader through displaying the result by figures.

The reminder of this paper is organized as follows. After presenting the utilized system model in section II, the optimization operation carried on by the GA is presented in section III. In section IV, are presented the derivation of the alternative bit mapping proposed, while in section V the optimized constellations are presented together with the numerical results obtained through simulation and with validation in realistic scenario. Section VI concludes the paper.

## II. SYSTEM MODEL

In Figure 1, the model used for the optimization is introduced by schematically describing the various building blocks of the communication system.

Considering $M$ to be the alphabet size of the constellation, each symbol represents a stream $u(\tau)$ of $log_2M$ bits generated by a memoryless source and is put in parallel by the serial-to-parallel (S2P) block.

These symbols are then modulated by the APSK modulation block giving place to a complex number that represents constellation symbols (1):

$$v_{n,m}(\tau) = \rho_n \cdot e^{i\theta_m} \quad (1)$$

where $\rho_n$ is the radius of the *n-th* circle of the constellation and $\theta_m$ is the phase of the *m-th* symbol. Figure 2 represents two examples for the 16 and 32-APSK.

Once constellation symbols are modulated through a squared root raised cosine (SRRC) filter, they are amplified by the high power amplifier (HPA) prior to transmission, thus being subject to the non-linear behaviour of the amplifier, whose effects can be modeled using the Saleh model [23] resulting in the output $s_m(\tau)$. This model distinguishes two effects:

- the AM/AM non-linear effect that models amplitude distortions on the input signal;
- the AM/PM non-linear effect that models phase distortions on the input signal.

This impairment can be efficiently reduced by ad-hoc pre-compensation at the transmitter. Despite its hardware complexity impact, commercial satellite modems have already adopted advanced dynamic pre-compensation techniques for standardized 16- and 32- APSK modes [24]. The dynamic pre-distortion algorithm takes into account the memory of the channel, conditioning the pre-distorted modulator constellation not only to the current symbol transmitted but also to the (L-1)/2 preceding and (L-1)/2 following symbols (L being the number of symbols in total) [25].

Considering the channel to be of the additive white Gaussian noise (AWGN) type, transmitted symbols are affected by the addition of a random nuisance signal with zero mean and variance $N_0/2$. Therefore, the received signal at the destination is $r(\tau) = S_m(\tau)+n(\tau)$. At the receiver, the signal is first passed through the SRRC filter. Note that due to the non-linear characteristics of the channel, this filter is no longer matched, and therefore ISI is introduced [23]. ISI appears at the receiver as the HPA, although memoryless, is driven by a signal with controlled ISI due to the presence of the modulator SRRC filter. This leads to an overall nonlinear channel with memory. As a consequence, the demodulator SRRC is not matched anymore to the incoming signal.

The resulting signal is passed to the demodulator that applies the maximum likelihood (ML) criterion in order to estimate the received symbol and it is then converted back from parallel to the serial signal $\hat{z}$.

In this work, to simplify the optimization at first the effect of SRRC filtering and therefore the ISI has been neglected. Moreover, only the AM/AM non-linear amplitude distortion $A(\rho)$, which represents the memoryless effect of the non-linearity, is taken into account through the formula

$$A(\rho) = \frac{\alpha \, \rho_n}{[1+\beta]\rho_n^2} \quad (2)$$

where $\alpha = 2.1587$ and $\beta = 1.1517$ are standard values of the constants gathered from the literature [26] and obtained by means of curve-fitting techniques. In section VI the above assumptions will be validated by testing the resulting optimized constellations in realistic scenario, that is considering real SRRC filters and opportune dynamic pre compensation techniques to mitigate ISI [11]. Dynamic symbol level pre-distortion considers the impact of memory introduced by the channel in order to compensate for the ISI.

The difference between the symbol generated by the memoryless source and the one estimated at the receiver can be computed as

$$d(\tau) = \left\| z(\tau) - \hat{z}(\tau) \right\| \quad (3)$$

and it is called *distortion*. The aim of the optimization presented in the next section is to minimize the average distortion defined as the minimum square error

$$MSE = E\left\{ [d(\tau)]^2 \right\} \quad (4)$$

which represents the optimization criterion. From (3) and (4) it can be noticed that errors in MSBs produce higher distortion, thus, higher MSE with respect to errors occurring in LSBs.

### III. GENETIC ALGORITHMS OPTIMIZATION

OP is tackled by means of GA, a well-known algorithm used in a variety of fields for all those non-convex problems that require a technique able to tentatively crawl the solution space in order to find the best one.

At each iteration $n$, a GA gives birth to a generation $G_n$ of potential solution vectors (also called chromosomes $\gamma_i$) that constitute the population of size $p$ of the OP:

$$G_n = \left\{ \gamma_1^{(n)}, \gamma_2^{(n)}, \ldots, \gamma_i^{(n)}, \ldots, \gamma_p^{(n)} \right\} \quad (5)$$

A vector of fitness scores $S_n$ is also calculated for each generation using the objective function $R$:

$$S_n = \left\{ R(\gamma_1^{(n)}), R(\gamma_2^{(n)}), \ldots, R(\gamma_i^{(n)}), \ldots, R(\gamma_p^{(n)}) \right\} \quad (6)$$

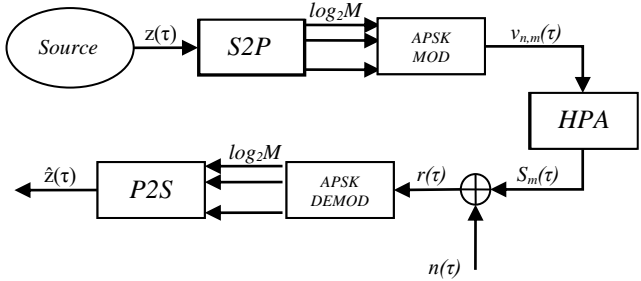

Figure 1. System model

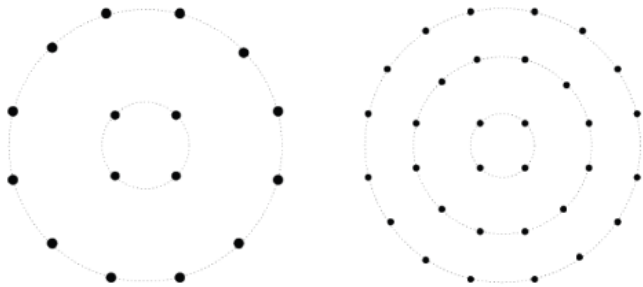

Figure 2. 16-APSK constellation (left) and 32-APSK constellation (right)

The chromosomes with the highest fitness score are meant to be the closest to the desired solution and are thus selected for surviving and giving place to the next generation.

The next generation is created in three steps:
1) *Selection* of the part of population with the best fitness score that will be parents for the next generation;
2) *Crossover* of selected parents according to a mixing criterion in order to give birth to a number of children from each couple that will constitute the next generation;
3) *Mutation* of a percentage of the offspring in order to spread the optimum solution search and avoid local optimal solutions.

The computation is stopped when the population has converged to the same fitness value which is supposed to be the optimal solution.

In the case of APSK the chromosomes are the radii and the phase of each symbol. The optimization criterion chosen is the minimization of the MSE, i.e., the expected minimum squared error between the transmitted and the received data. Therefore, the function used to calculate the fitness scores is *R=1/(MSE)*. The optimization is constrained by the assumption that the peak-power limit imposed over the constellation cannot exceed the saturated power of the HPA.

For a more in-depth treatise about the use of GA in the context of constellation optimization, the reader can refer to [15] and [16].

### IV. MAPPING ANALYSIS

The goal of this paper is to demonstrate that the performance of 64-APSK modulation can be improved in two successive steps: at first choosing opportune alternative mappings with respect to conventional CCSDS ones and then optimizing the position of the constellation symbols in terms of amplitude and phase, accordingly. These two options are strictly related, since in the optimization step the GA could modify the position (i.e., radius and angle) of the symbols in such a way that symbols on the same ring result interchanged thus modifying the mapping. In this section the influence of the mapping in term of inter symbol distortion will be analyzed. Consider the constellation described as $a+b+c+d$, where $a$ is the number of symbols in the inner circle, $d$ is the number of symbols in the outer circle and so forth. The transmission of equiprobable symbols using the 64-APSK scheme proposed by CCSDS [20] (that is of the type 4+12+20+28) is first presented. Afterwards, a new mapping of the 4+12+20+28 type and one of the 4+12+16+32 type specifically designed for the case under investigation are presented and compared with the CCSDS one.

#### A. CCSDS Mapping for the 4+12+20+28 constellation

As previously said, the reference mapping used for the 4+12+20+28 constellation is the CCSDS mapping [20]. The considered mapping is symmetric with respect to the *x* and *y* axes. The advantage of such a mapping resides in its simplicity, since the coordinates of the symbols from a single quadrant are sufficient to describe the entire constellation. Therefore, in this case, the chromosomes used for the successive optimization will be as follows:

$$\gamma = [\rho_0, \rho_1, \rho_2, \theta_0, \theta_1, \theta_2, \theta_3, \theta_4, \theta_5, \theta_6, \\ \theta_7, \theta_8, \theta_9, \theta_{10}, \theta_{11}, \theta_{12}, \theta_{13}, \theta_{14}, \theta_{15}] \quad (6)$$

with $\rho_0, \rho_1, \rho_2$ indicating the three inner radii (the outer one is always equal to one thus not taking part in the optimization phase), and values from $\theta_0$ to $\theta_{15}$ indicating the angles for the symbols with value from 0 to 15. Although symmetric mapping reduces the number of variables in the optimization problem, this limitation yields to a bigger distortion. In fact, in Figure 3(a) it is noticeable that the regions with a higher distortion value are those close to both the horizontal and the vertical axis. The meaning of numbers between the connection in each constellation figures represents such distortion. To reduce this effect, it is necessary to eliminate those critical zones dropping some of the symmetry constraints thus adding more variables and more complexity to the optimization problem. Considering that solving the optimization problem using 64 independent symbols is computationally impractical, exploiting the symmetry the number of variables involved is equal to 35.

### B. Proposed Mapping for the 4+12+20+28 constellation

The presented mapping has only a vertical symmetry and 35 variables: 3 regarding the radii and 32 regarding the angles. Exploiting the symmetry drop it is possible to radically change the mapping so that the distortion of the symbols of the outer radius that are the closest to the horizontal axis is minimized.

The resulting mapping is represented in Figure 3(b), where it is noticeable that symbols near the vertical axis keep high distortion levels as a consequence of the symmetry choices made in the design phase. In the horizontal axis instead, the distortion among symbols varies from 1 to 10, while in the case described in the previous section it was constantly equal to 16. This variety of distortion zones will be beneficial in the constellation optimization phase, since it will allow to keep higher distance between symbols which potentially can cause higher levels of distortion.

In Table I it can be seen that the average distortion for symbols on the same ring is smaller than in the CCSDS mapping case. Another important aspect is the distortion among symbols on different rings. This parameter is obtained averaging the distortion among all the neighbor symbols on two rings.

### C. Proposed Mapping for the 4+12+16+32 constellation

Sometimes, as shown in Figure 3(c), the 64-APSK constellation operates with 4 more symbols in the outer ring. This yields to smaller regions of decision in the outer part of the constellation, but leaves more space in the inner circles so that distortion can be more efficiently reduced in the inner zone. As in the case treated in section *IV.B*, also here only the vertical symmetry has been kept and the intra-ring distortion has been reduced especially in the outer symbols.

Table II shows the distortion among symbols in the same ring. As it can be seen, moving 4 symbols to the outer radius the resulting distortion is reduced while the one on the third ring is increased. This allows to better optimize the constellation through the GA, since now the third ring has more space to put more distance between those symbols that create critical levels of distortion. In this case, comparing inter-ring distortion would lose its meaning, since the number of symbols between the two outer rings is changed.

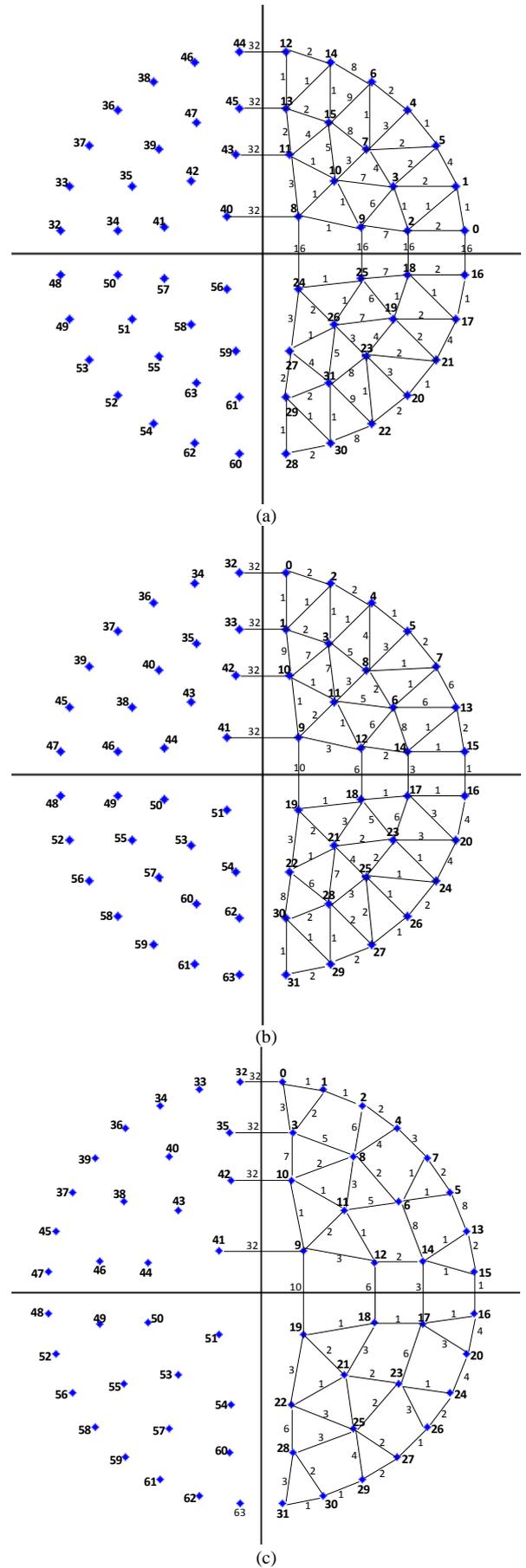

Figure 3. Distortion of the CCSDS mapping for the 4+12+20+28 (a), the 4+12+20+28 (b) and the 4+12+16+32 composed 64-APSK constellation (c). Symbols and relative mutual distortion are expressed in decimal representation.

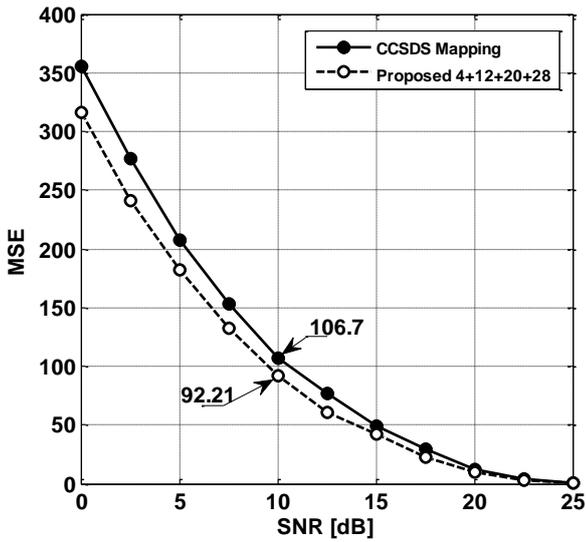

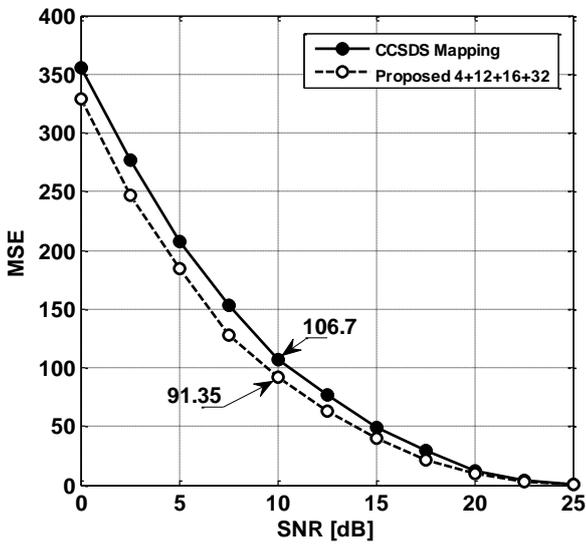

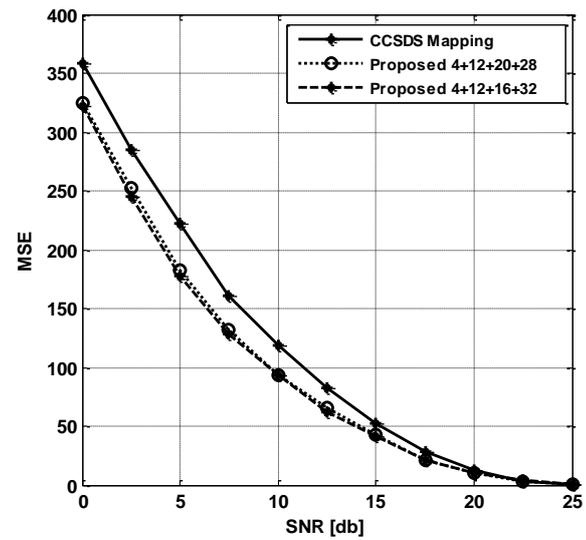

Figure 4. MSE results for the CCSDS mapping and the 4+12+20+28 (a), 4+12+16+32 (b) and overall comparison of the presented constellations without GA optimization, (c).

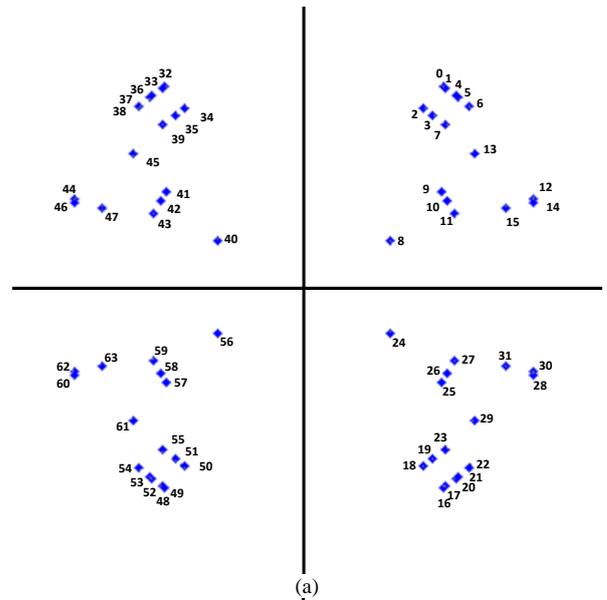

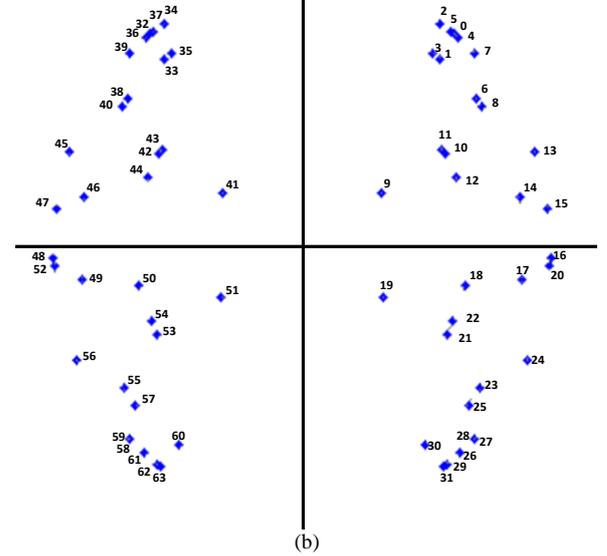

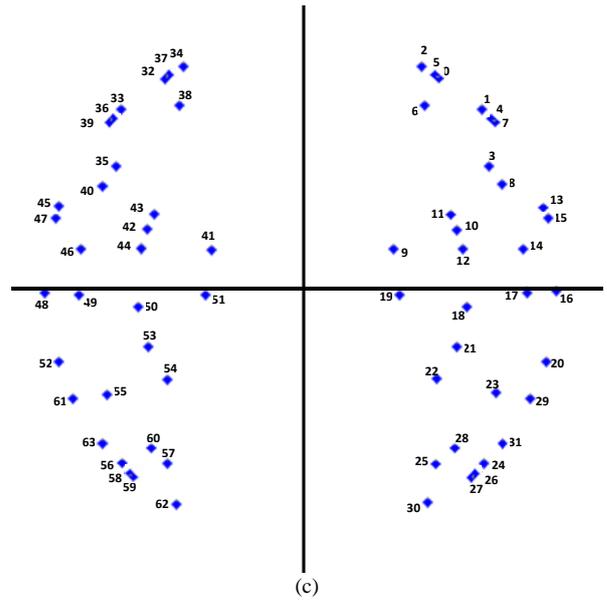

Figure 5. Optimized constellation for the 4+12+20+28 CCSDS mapping (a), 4+12+20+28 proposed mapping (b), and 4+12+16+32 proposed mapping (c). Symbols are expressed in decimal representation.

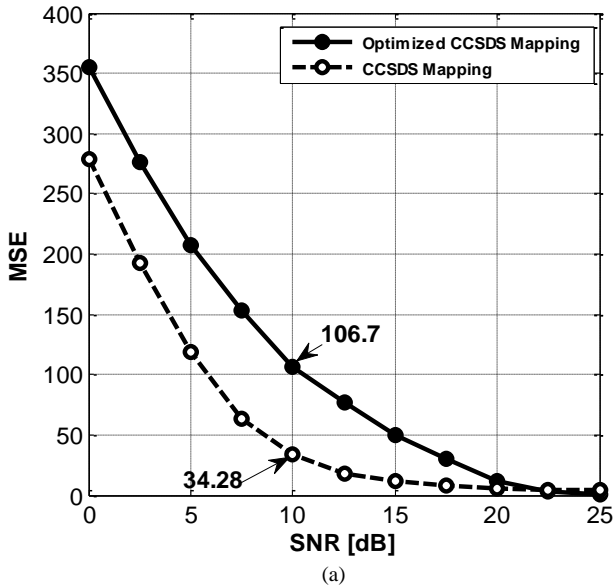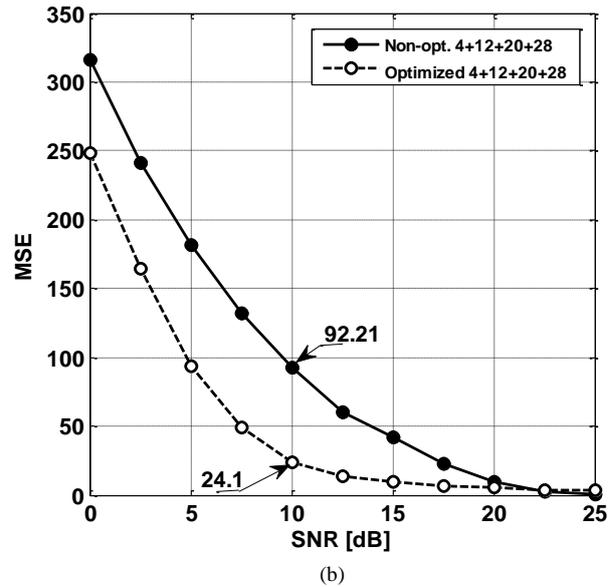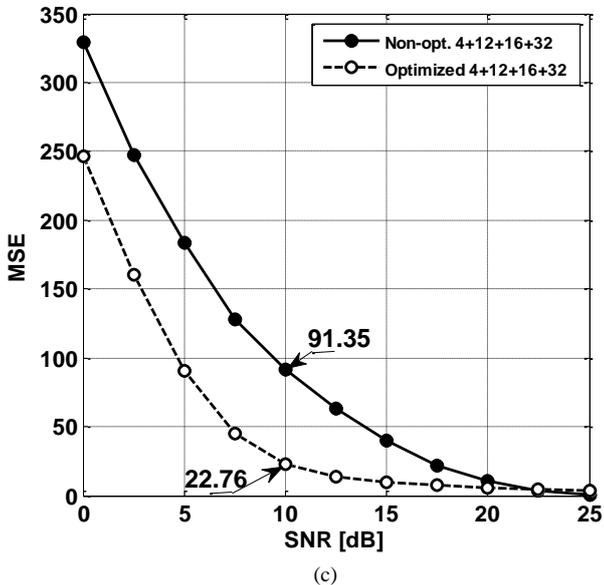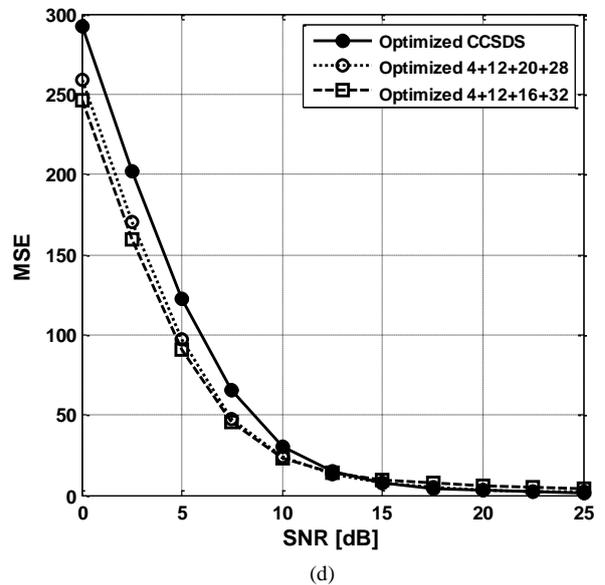

Figure 6. Performance comparison between optimized and non-optimized CCSDS mapping (a), performance comparison between optimized and non-optimized 4+12+20+28 proposed mappings (b), MSE results obtained optimizing the CCSDS and the 4+12+16+32 proposed mappings (c), and overall comparison of the presented constellation with GA optimization (d).

TABLE I
COMPARISON OF THE AVERAGE DISTORTION AMONG NEIGHBOR SYMBOLS ON DIFFERENT RINGS

|  | CCSDS Mapping 4+12+20+28 | Proposed Mapping 4+12+20+28 |
|---|---|---|
| Outer ring vs. Ring 3 | 2.27 | 1.77 |
| Ring 3 vs. Ring 2 | 4.85 | 5.42 |
| Ring 2 vs. Inner ring | 2 | 2 |

TABLE II
COMPARISON OF THE AVERAGE DISTORTION AMONG NEIGHBOR SYMBOLS ON THE SAME RING

|  | CCSDS Mapping 4+12+20+28 | Proposed Mapping 4+12+20+28 | Proposed Mapping 4+12+16+32 |
|---|---|---|---|
| Outer ring | 6 | 4.5 | 4.18 |
| Ring 3 | 7.8 | 6.5 | 7.65 |
| Ring 2 | 8.66 | 7.33 | 7.16 |
| Inner ring | 24 | 21 | 21 |

## V. OPTIMIZATION RESULTS

### A. Results without constellation optimization

In the case of CCSDS presented in Figure 4(a), the first thing to notice is that the *MSE* = 106.7 at *SNR* = 10 *dB* while the average distortion is 10.32, that is approximately 16% of the maximum distortion level for 64-APSK. As already stated, the reason of this high level of distortion can be found in the presence of critical zones close to the axes. In the case of 28 symbols in the outer ring and proposed mapping, the mitigation of those high distortion zones already yields to better results for conventional uniform constellation. In particular, when *SNR* = 10 *dB* the *MSE* = 92.21 that is 13.6% less than in the CCSDS mapping case. In Figure 4(b), the distortion in the outer ring for *SNR* = 10 *dB* of the proposed constellation with 32 symbols, is 91.35 that means a reduction of 14.3% with respect to the CCSDS constellation.

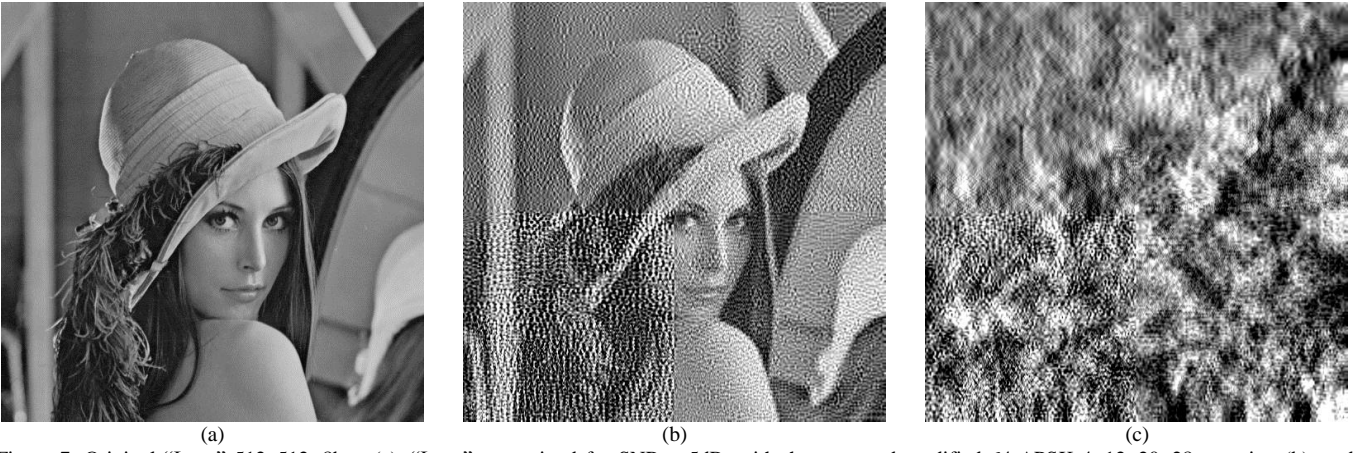

(a) (b) (c)

Figure 7. Original "Lena" 512x512, 8bpp (a), "Lena" transmitted for SNR = 5dB, with the proposed modified 64-APSK 4+12+20+28 mapping (b), and conventional 64-APSK with CCSDS mapping (c).

Finally, in Figure 4(c) a comparison among the three mappings is given demonstrating that regardless of the number of symbols in the outer rings, the two proposed mappings have similar performances, with a slight predominance of the 4+12+16+32 case.

### B. Results with constellation optimization

At first, to set up the GA evolution we refer to a previous study performed in [16] which analyzed the influence of the GA parameters on the accuracy and convergence performance of the algorithm. Based on [16] only 10 % of a chromosome could vary and a maximum of 40% of the chromosome parents are allowed to appear on the next generation. In the results that follow, the GA is run with a population of 100 chromosomes characterized by genes with values uniformly distributed on the constellation circles and probabilities of *mutation* and *crossover* $P_{mut} = 0.3$ and $P_{cross} = 0.5$, respectively.

Using these values, transmission over satellite is simulated according to the model presented in Section II. Then, the fitness function is evaluated thanks to the function $R$ already discussed. Once this step is completed, the GA modifies the population in accordance with the fitness results using the policies defined by the specific *selection* and *crossover* functions taken into consideration, that in this case are the *tournament* selection function and the *two-point* crossover function, chosen using the procedure already presented in [16]. The transmission is then repeated using the new generation until convergence or the maximum number of generations (in this case set to $n = 200$) is reached.

The considered chromosomes are respectively

- $\gamma = [\rho_0, \rho_1, \rho_2, \theta_0, \theta_1, \theta_2, \theta_3, \theta_4, \theta_5, \theta_6, \theta_7,$
  $\theta_8, \theta_9, \theta_{10}, \theta_{11}, \theta_{12}, \theta_{13}, \theta_{14}, \theta_{15}]$

for the case of the CCSDS mapping;

- $\gamma = [\rho_0, \rho_1, \rho_2, \theta_0, \theta_1, \theta_2, \theta_3, \theta_4, \theta_5, \theta_6, \theta_7,$
  $\theta_8, \theta_9, \theta_{10}, \theta_{11}, \theta_{12}, \theta_{13}, \theta_{14}, \theta_{15}, \theta_{16}, \theta_{17}, \theta_{18}, \theta_{19},$
  $\theta_{20}, \theta_{21}, \theta_{22}, \theta_{23}, \theta_{24}, \theta_{25}, \theta_{26}, \theta_{27}, \theta_{28}, \theta_{29}, \theta_{30}, \theta_{31}]$

for the proposed new mapping.

TABLE III
COMPARISON OF THE PEAQ USING NON-LINEAR AND IDEAL HPA FILTER

|  | CCSDS Mapping 4+12+20+28 | Proposed Mapping 4+12+20+28 |
|---|---|---|
| Non-linear HPA | -3.47 | -3.34 |
| Ideal HPA | -3.57 | -3.37 |

The three radii indicate the three inner layers (since the outer one is normalized to 1, while the phases have been defined so that the subscripts of each theta corresponds to the alphabet value assigned to that symbol.

The optimized constellation using the mapping proposed by CCSDS and described in section *IV.A* is shown in Figure 5(a). Figure 5(b) shows the optimized constellation for the optimized 4+12+20+28 mapping proposed in section *IV.B* while in Figure 5(c) the optimized 4+12+20+32 proposed constellation is drawn.

In Figure 6(a), for *SNR = 10 dB* the *MSE = 34.28* with a reduction of 67% with regard to the case of non–optimized constellation described in Figure 4(a). The optimization is more evident for low SNR while the two curves converge for higher values.

In Figure 6(b), a comparative analysis at *SNR = 10 dB* shows that for optimized 4+12+20+28 mapping the MSE =24.1 that is 77% less with regard to the non-optimized case. Similarly, in Figure 6(c), when *SNR = 10 dB* the average distortion for optimized 4+12+16+32 mapping is 22.76 with a reduction of the 78% with regard to the non-optimized mapping.

Figure 6(d) puts compares the results for the 3 constellations optimized through GA presented in figures 5(a), (b), and (c), showing once again that placing 28 or 32 symbols on the outer ring yields to similar results.

### C. Validation in realistic scenario

In this section the performance of the proposed optimized constellations over a realistic system model is investigated. The roll-off factor of the SRRC filter is fixed to 0.2 and dynamic pre-distortion have been considered using *L = 3* symbols simultaneously for reducing the ISI introduced by the non-linear HPA [11].

Furthermore, to assess the suitability of the proposed approach with real applications. At first a 5 seconds from a stereo audio CD signal sampled at 44.1 KHz coded at 16 bps has been considered. The measurements have been made using the computer measuring system Opera™ comparing them with the referential clip. Again for SNR = 10 *dB*, the PEAQ measured on the received audio sequence transmitted with the proposed optimized 64-APSK constellation with proposed modified 4+12+20+28 mapping is -3.34, against the conventional 64-APSK with CCSDS mapping which achieved a lower score of -3.47. The obtained PEAQ score is just below the typical performance of low bit rate audio codecs: for the transmission of audio sequences at a rate of 64 Kbit/s, MP3/MPEG-4 codecs achieve PEAQ of around -3.36 [22]. As detailed in Table III, notice that the same transmission performed with ideal HPA under the simplified hypothesis specified in section II and used in section *V.A* and *V.B* produced a score of -3-37 for the proposed optimized 64-APSK constellation with proposed modified 4+12+20+28 mapping and a PEAQ of -3.57 for conventional 64-APSK with CCSDS mapping. In fact, this shows that dynamic pre-distortion can outperform the ideal HPA since dynamic pre-distortion also reduce ISI while the ideal HPA is a memoryless device that cannot reduce ISI. It is worth noting also that the gain obtained when dynamic pre-compensation is used is higher for conventional 64-APSK with CCSDS mapping against the proposed optimized 64-APSK constellation with proposed modified 4+12+20+28 mapping. Indeed, in the proposed optimized constellation the symbols tend to be approaching the outer ring, thus being more likely to cause performance degradation due to ISI.

To further assess the correspondence between MSE and the subjective perception the transmission of a standard test grey-level image "Lena" of size 512x512 coded at 8 bpp has been considered. Figure 7 shows the original image and the images received in case of a transmission for SNR = 5 *dB*. Figure 7(b) is obtained by adopting the proposed optimized 64-APSK constellation with proposed modified 4+12+20+28 mapping, whereas figure 7(c) is obtained by using the conventional 64-APSK with CCSDS mapping. It can be observed that in the second case the image is completely incomprehensible, whilst in the first case it can be easily understood in spite of the errors due to the highly noisy channel.

## VI. CONCLUSIONS

In this paper, the use of modified constellations and mapping for transmission in the satellite broadcasting scenario has been proposed for 64-APSK to achieve unequal error protection of data which exhibits different sensitivity to channel errors (e.g., multimedia signals). In particular, new mappings derivation starting from conventional CCSDS mapping together with constellation optimization using genetic algorithms have been presented. Based on the assumption that errors in most significant bits (MSB) produce higher distortion than errors in the least significant bits (LSB), the proposed optimization aggregates symbols that have lower inter-symbol distortion and separates symbols which have higher inter-symbol distortion. Found results demonstrate that it is possible to improve the performance of conventional 64-APSK in terms of average distortion between transmitted and received data (i.e., MSE) using modified mappings where some symmetries are dropped and the symbols on the same circles are not equally spaced anymore, thus producing asymmetric constellations. Tests using real multimedia signals in realistic transmission scenario allow assessing the suitability of the proposed scheme for digital multimedia broadcasting.

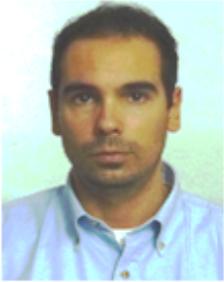

**Matteo Anedda** received the B.Sc. degree in Electronic Engineering in 2011 and the M.Sc degree (Summa cum Laude) in Telecommunication Engineering in 2012, both from the University of Cagliari. He is currently pursuing the Ph.D. degree in the Multimedia Communication Lab (MCLab), Department of Electrical and Electronic Engineering (DIEE), University of Cagliari. He has been a visiting student at the University of Basque Country (EHU/UPV), Bilbao, Spain, in 2010, where he has carried out his M.Sc. thesis with Prof. Pablo Angueira. He received an award for his Master thesis from Order of the Engineers of Cagliari. His research interests are network planning, coverage optimization and optimization algorithms. Dr. Anedda is a member of the IEEE Membership, IEEE Communications Society, IEEE Young Professionals, IEEE Broadcast Technology Society Membership and IEEE Vehicular Technology Society Membership.

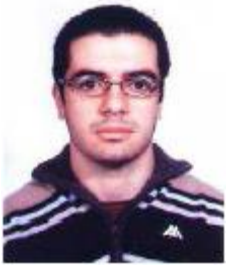

**Alessio Meloni** received his M.Sc. degree (Summa cum Laude) in electronic engineering from the University of Cagliari (Italy) in December 2010, after spending one year at the Czech Technical University (Prague) during which he developed his thesis. Since March 2011, he has been with the Multimedia and communications Lab of the Department of Electrical and Electronic Engineering (DIEE) - University of Cagliari, where he gained his Ph.D. degree on Advanced Random Access techniques for satellite communications in March 2014. From April 2011 to October 2011 he has been guest PhD student at the German Aerospace Center, Oberpfaffenhofen, Germany. His research interests include wireless communications, advanced random access techniques and satellite links. He is a member of IEEE, IEEE ComSoc, CNIT and GOLD member of BTS.

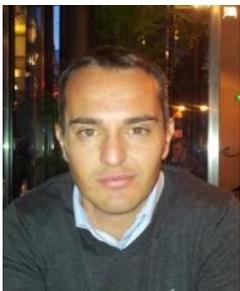

**Maurizio Murroni** (M'01) (SM'13) is Assistant Professor of Communication at the Department of Electrical and Electronic Engineering (DIEE) of the University of Cagliari. He graduated (M.Sc.) with honors (Summa cum Laude) in Electronic Engineering in 1998 at the University of Cagliari. In 2001 he received his Ph.D. degree in Electronic Engineering and Computers, from the University of Cagliari. Since the 1998, he contributes to the research and teaching activities of the Multimedia Communication Laboratory (MCLab) at DIEE. He was graduate visiting scholar at CVSSP Group, School of Electronic Engineering, Information Technology and Mathematics, University of Surrey, Guildford, U.K. in 1998 and a visiting Ph.D. scholar at the Image ProcessingGroup, Polytechnic University, Brooklyn, NY, USA, in 2000. In 2006 he was visiting lecturer at the Dept. of Electronics and Computers at the Transilvania University of Brasov in Romania and in 2011 visiting professor at the Dept. Electronics and Telecommunications, Bilbao Faculty of Engineering, University of the Basque Country (UPV/EHU) in Spain. Since October 2010 he is coordinator of the research unit of the Italian University Consortium for Telecommunications (CNIT) at the University of Cagliari. Dr. Murroni is author of more than 50 journal articles and peer-reviewed conference papers. He served as a technical program chair for various international conferences and workshops and as a reviewer and panelist for several funding agencies, including Italian MIUR and EU Regional funding agency. His research has been funded by European, national, regional and industrial grants and currently focuses on Broadcasting, Cognitive Radio system, signal processing for radio communications, multimedia data transmission and processing. Dr. Murroni is a Senior member of IEEE, IEEE BTS, IEEE Com Soc, IEEE DySPAN-SC and 1900.6 WG.